\documentclass[doublecol]{epl2} 

\title{Spatial repartition of local plastic processes in different
  creep regimes in a granular material} \shorttitle{Title}

\author{A. Pons\inst{1} \and T. Darnige\inst{1} \and
  J. Crassous\inst{2} \and E. Cl\'ement\inst{1} \and A. Amon\inst{2}}

\shortauthor{F. Author \etal}

\institute{ \inst{1} Physique et M\'ecanique des Milieux
  H\'et\'erog\`enes, PMMH ESPCI, CNRS (UMR7636) - Univ. P.M. Curie -
  Univ. Paris-Diderot - PSL Research University - 75005 Paris,
  France\\ \inst{2} Institut de Physique de Rennes (UMR URI-CNRS
  6251), Univ. Rennes 1, Campus de Beaulieu, Rennes, France }

\abstract{Granular packings under constant shear stress display below
  the Coulomb limit, a logarithmic creep dynamics. However the
  addition of small stress modulations induces a linear creep regime
  characterized by an effective viscous response. Using Diffusing Wave
  Spectroscopy, we investigate the relation between creep and local
  plastic events spatial distribution (``hot-spots'') contributing to
  the plastic yield. The study is done in the two regimes, i.e. with
  and without mechanical activation. The hot-spot dynamics is related
  to the material effective fluidity. We show that far from the
  threshold, a local visco-elastic rheology coupled to an ageing of
  the fluidity parameter, is able to render the essential
  spatio-temporal features of the observed creep dynamics.}
	

\begin{document}

\maketitle

\section{Introduction}

Granular packings are often seen as rigid below a limit corresponding
to a critical ratio between shear stress and normal stress (Coulomb
threshold) \cite{Wood1990}. However, the existence of a clear-cut
transition between a solid-like and liquid-like behaviour is currently
strongly challenged \cite{Nichol2010,
  Reddy2011,Nguyen2011,Amon2012}. In the presence of a shear band
(i.e. a fluid zone dwelling somewhere in the packing) different
authors brought evidences for mechanically activated creeping
processes taking place in remote regions, below the Coulomb
threshold\cite{Nichol2010, Reddy2011}. This behaviour led to non-local
rheological relations proposed to extend the standard local
constitutive relations for granular flows \cite{Henann2013,
  Bouzid2013,Henann2014, Bouzid2015}. For granular packing sheared in
all of its parts below the Coulomb limit, dynamical processes leading
to a logarithmic creep occurs \cite{Schmertmann1991,
  Nguyen2011,Amon2012}. Interestingly, this creeping dynamics can be
mapped onto a simple visco-elastic model initially designed to render
the phenomenology of yield stress fluids displaying ageing in the
solid phase \cite{Derec2001}. The model is centred on a dynamical
equation for a fluidity parameter representing an effective
visco-elastic relaxation. This phenomenological parameter was directly
related to the occurrence of mesoscopic plastic events called ''hot
spots'' \cite{Amon2012}. In the vicinity of the dynamical threshold,
these events combine to provide large scale plastic yields
\cite{Amon2012, LeBouil2014}. Recently, we have shown that providing a
tiny stress modulation around a nominal shear stress, the creep
dynamics changes from a logarithmic to a linear behaviour
\cite{Pons2015}. The physical interpretation stems from the
combination of memory effects and non-linearities, leading to a
''secular'' accumulation of tiny effects, meaning that the creep
dynamics is revealed at a time scale much larger than the
modulation. We call this behaviour \emph{``rectified creep''} in the
following as the interpretation of this regime is different from an
Eyring-like activated process. As the ingredients at the origin of
this rectified creep are generic for a large class of soft glassy
materials, this effect should be seen in other yield stress fluids
displaying
creep~\cite{Moller2009,Marchal2009,Negi2010,Siebenburger2012}.

In this paper, using a spatially resolved multiple scattering
technique \cite{Erpelding2008}, we monitor the spatial distribution of
hot-spots during the two creep regimes. In parallel, a spatially
resolved visco-elastic model is solved assuming a direct relation
between the hot-spots production rate and the local fluidity
value. The numerical solution is compared to the
experiments. Interpretation of the results in the framework of a local
rheological model is provided.


\begin{figure*}[htbp]
\centerline{\includegraphics[width=0.95\linewidth]{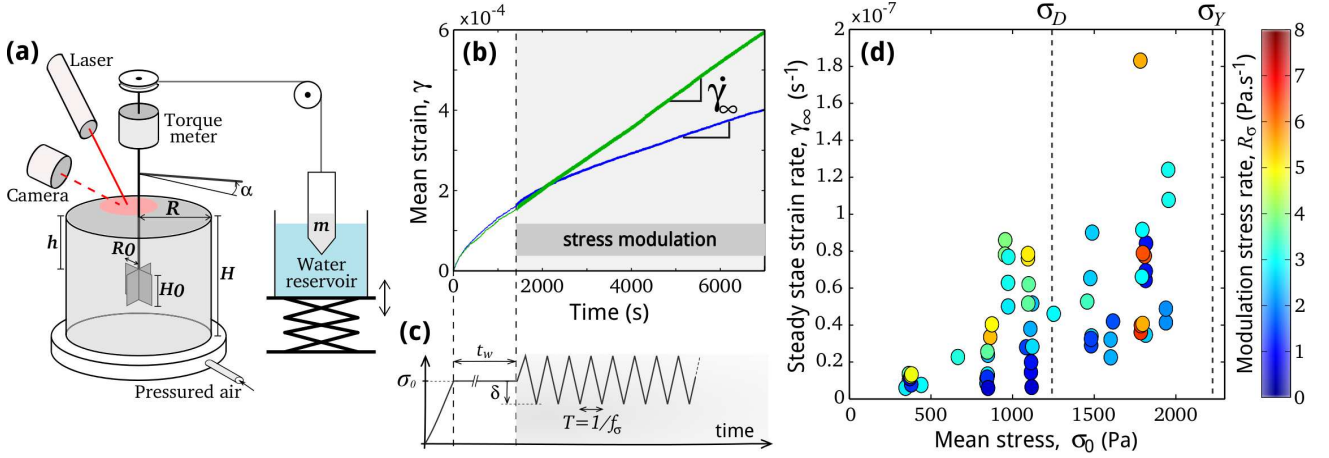}}
\caption{ (a) Experimental set-up: shear is imposed by an hanging mass
  $m$ connected to the vertical axis; vertical displacement of the
  water reservoir allows modulation of the applied stress; a camera
  collects backscattered light from a He-Ne laser illuminating the top
  of the shear cell. (b) Mean strain as a function of time for two
  experiments performed at $\sigma_0 = 1100 Pa$, and $\delta = 7.5Pa$
  , and for two oscillations frequencies: 0.099 Hz (blue line) and
  0.168 Hz (green line). The oscillations start at $t=1500s$ (grey
  area). $\dot{\gamma}_\infty $ is the slope of the linear part. (c)
  Imposed stress during an experiment: stress ramp to reach mean
  stress, $\sigma_0$, constant stress during $t_w$ and stress
  modulation characterized by a frequency $f_\sigma$ and an amplitude
  $\delta$. (d) Steady state strain rate, $\dot{\gamma}_\infty $, as a
  function of mean stress, $\sigma_0$. Color of the symbol stands for
  the value of modulation stress rate $R_\sigma=4\delta f$. }
\label{fig.setup}
\end{figure*}
\section{Methods}
The experimental set-up is shown in fig.~\ref{fig.setup}a.  It
consists in a cylindrical shear cell (Radius $R=5cm$, height $H=10cm$)
filled with glass beads of density $\rho=2500\ kg/m^3$ and mean
diameter $d=200\ \mu m$ (rms polydispersity $\Delta d= 30\ \mu m$). A
well defined packing fraction $\Phi = 0.605\pm0.005$ is obtaining
using an air fluidized bed as described by Nguyen et
al. \cite{Nguyen2011}.

Shear is obtained by applying a torque on a stainless steel four-blade
vane (radius $R_0 =1.27\ cm$, height $H_0 =2.54\ cm$) via a mass $m$
suspended from a pulley (see fig.~\ref{fig.setup}a). Using this
``Atwood-machine" technique, mechanical noise inherently coming from
any motorized process can be suppressed.  The mass $m$ hangs partially
inside a reservoir filled with water. Thus, by modulating the
Archimedes force, through the up and down motion of the reservoir
sitting on a vertical translation stage, controlled modulation of the
torque applied to the granular packing can be obtained.
A torque probe connected to the vane axis measures the applied torque
$T$. The vane rotation angle $\alpha$ is monitored via a transverse
arm whose displacement is measured by an induction probe. Torque and
displacement signals as well as the vertical translation stage command
are connected to a Labview controller board. We defined here the mean
stress and the mean strain as $\sigma_0=\frac{T}{2\pi R_0^2H_0}$ and
$\gamma=\frac{\alpha R_0}{R-R_0}$ respectively.

For a given experiment, the protocol (fig.~\ref{fig.setup}c) is the
following $(i)$ stress ramp at constant stress rate ($\dot{\sigma} =5
\ Pa/s$) up to the desired mean stress value $\sigma_0$; $(ii)$
constant shear $\sigma_0$ applied during $t_w=1500\ s$; $(iii)$
modulation of the stress around $\sigma_0$ for at least 2 hours. The
modulation consist in triangular oscillations with an amplitude
$\delta$ and a frequency $f_\sigma$. We introduce the modulation
stress rate, $R_\sigma=4\delta f_\sigma$ to characterize the
modulation. 

Let us note that we used the same set-up and protocol than
\cite{Pons2015} and results presented here include those already
presented. The two following points differs from the former study.

Here, two vane penetration depths were used : $h=5\ cm$ and
$0\ cm$. During the vane insertion procedure, in order to prevent
large scale disturbances in the packing, pressurized air is gently
flown, just below the fluidization threshold providing a packing at
the surface bearing almost no confining pressure. In addition, after
the introduction of the vane, we kept the air flowing during 10 min in
order to relax the remaining stress perturbations induced by the vane
insertion. The air flow is switched off before the start of the
experiment. For experiments performed with the vane near the surface
($h=0\ cm$), the grains are confined with a glass plate in order to
obtain a confining pressure at the top of the vane close to the one
existing for experiments done at the insertion height $h=5\ cm$. The
circular glass plate has a central hole to let the vane shaft go
through and the vertical confinement pressure was adjusted by placing
loads on the glass plate.

In addition to the mechanical measurements, we obtain, by diffusive
wave spectroscopy (DWS) \cite{Erpelding2008,Erpelding2010,Amon2012}, a
spatially resolved map of the top surface deformations. DWS is an
interference technique using scattering of coherent light by strongly
diffusive materials. In our case, a He-Ne laser ($\lambda=633\ nm$)
illuminates the top of the shear cell. A camera imaging the surface at
a frame rate of 0.1 Hz collects back-scattered light
(fig.~\ref{fig.setup}a). The correlation of scattered intensities
between two successive images, $g_I$, is computed by zones of 16
$\times$ 16 pixels, composing correlation maps of 370 $\mu m$
resolution.  The link between the value of $g_I$ and the corresponding
deformations that have occurred inside the sample were extensively
described in \cite{Erpelding2008,Erpelding2010}. Maximal correlation
($g_I>0.99$, white on Fig.~\ref{fig.DWS}a) corresponds to an
homogeneous deformation below $10^{-7}$ and vanishing correlation
(black on Fig.~\ref{fig.DWS}a) corresponds to deformations larger than
$10^{-5}$.


\section{Experimental results} 

Figure~\ref{fig.setup}.b shows typical deformations for two
experiments performed at the same mean stress ($\sigma_0 = 1100 Pa$)
and oscillation amplitude ($\delta = 7.5Pa$) for two oscillation
frequencies. These two experiments were performed at a penetration
depth $h=5cm$.  During the phase at constant stress, we observe a slow
increase of the deformation, $\gamma(t)$, similar for both
experiments. This initial dynamics can be fitted by a logarithmic
curve as in \cite{Nguyen2011}.  Then, when submitted to oscillations,
the system transits to a linear creep regime characterized by a
constant mean strain rate, $\dot{\gamma}_\infty $, which increases
with the oscillation frequency. This sub-threshold fluidization by
mechanical perturbation has been already reported and interpreted
in~\cite{Pons2015} as a secular drift
process. Figure~\ref{fig.setup}.d shows the value of the steady state
strain rate, $\dot{\gamma}_\infty $, for all the experiments performed
at $h=5cm$ and at various $\sigma_0$ and $R_\sigma=4\delta
f_\sigma$. Though the data are dispersed, $\dot{\gamma}_\infty $
clearly increases with $\sigma_0$ and $R_\sigma$ as predicted by
\cite{Pons2015} as long as $\sigma_0$ is below $\sim1500Pa$. For
higher stresses, $\dot{\gamma}_\infty $ keeps increasing with
$\sigma_0$ but the influence of $R_\sigma$ remains unclear. For these
stress values around the stress dynamical threshold (see
\cite{Nguyen2011}), the mean shear rate $\dot{\gamma}_\infty$ indeed
becomes very large, however the experiments display a great amount of
sensitivity to preparation, which makes measurements in this limit
quite difficult.

\begin{figure}[htbp]
\centerline{\includegraphics[width=0.95\linewidth]{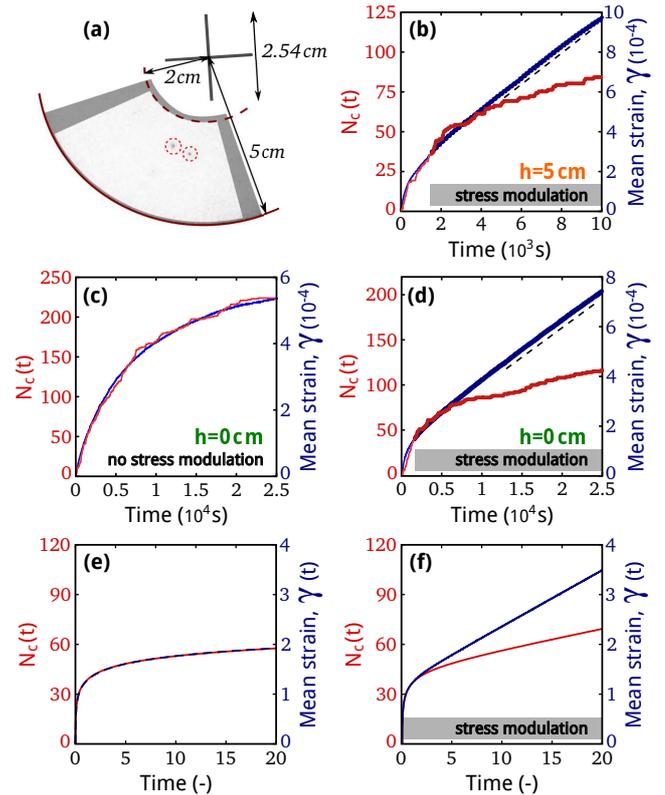}}
\caption{(a) Typical correlation map of the top surface. Two typical
  ``hot spots" are visible. The shaded part is not used in the
  analysis. (b-d) Comparison of the mean strain, $\gamma(t)$ (blue)
  and the cumulative number of spots, $N_{c}(t)$ (red) as a function
  of time for three experiments performed at the same mean stress,
  $\sigma_0=1000Pa$ without (b) or with (c-d) stress modulation
  ($\delta=10Pa$, $f_{\sigma}=0.1Hz$). Experiments (c) and (d) are
  performed at $h=0\ cm$, experiment (b) at $h=5\ cm$. The grey strip
  shows when the modulation is on. (e)-(f) Mean strain, $\gamma(t)$
  (blue) and cumulative fluidity, $N_{c}(t)$ (red) obtained by the
  numerical resolution of
  eqs.~(\ref{eq.fluidity1bis}-\ref{eq.fluidity2bis}) taking into
  account the Couette geometry of the system, without (e) and with (f)
  stress modulation around the same mean stress. Parameters : $a=1$,
  $\sigma_0 = 0.8\sigma_D$, $f_0=1$, modulation: $\delta =
  0.005\sigma_D$ and $f_{\sigma} = 0.2$.}
\label{fig.DWS}
\end{figure}

Several experiments performed at either $h=0$ or $5\ cm$ were coupled
to the DWS technique. Fig.~\ref{fig.DWS}a represents a typical map of
the top surface deformation: in average, the correlation, $g_I$, is
larger than 0.99 (white) implying a low and homogeneous deformation
except over small areas (black spots), called ``hot spots"
in~\cite{Amon2012}, which are characterized by relatively large
plastic deformations. Figures~\ref{fig.DWS}b-d show the mean strain,
$\gamma(t)$ and the cumulative number of ``hot spots", $N_{c}(t)$,
over the surface of interest, $S$, as a function of time for three
experiments performed at the same mean stress, $\sigma_0=1000Pa$. For
the experiment shown in Fig.~\ref{fig.DWS}c no modulation is performed
unlike the ones shown in Fig.~\ref{fig.DWS}b and d. Experiments b and
d differ in the depth of the vane, respectively $h=5\ cm$ and
$h=0\ cm$. One can note that the ``hot-spots" are present at the top
surface even when the vane is buried inside the packing
(fig.~\ref{fig.DWS}b) which indicates that the localized relaxation
process is spreading over the whole material.

\begin{figure}[h!]
\centerline{\includegraphics[width=0.95\linewidth]{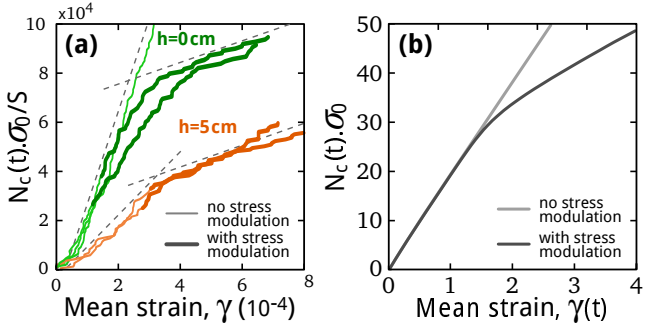}}
\caption{Normalized cumulative number of hot-spots as a function of
  the mean strain for 5 different experiments performed at
  $\sigma_0=1000Pa$. Green lines: $h=0cm$, orange lines: $h=5cm$. Fine
  line: no modulation, thick lines: with modulation. Dash lines are
  guide to the eye. (b) Corresponding graph obtained by numerical
  resolution of eq.~(\ref{eq.fluidity1bis}-\ref{eq.fluidity2bis})
  taking into account the geometry of the system. Same parameters as
  Fig~\ref{fig.DWS}.e-f.}
\label{fig.NG}
\end{figure}

Fig.~\ref{fig.NG}a shows $N_{c}(t)$ normalized by $S/\sigma_0$ as a
function the global plastic deformation for experiments performed at
the same stress for either $h=0\ cm$ (green) or $h=5\ cm$ (orange).
Without modulation (fine lines in Fig.~\ref{fig.NG}.a), the
proportionality between $N_{c}(t)$ and $\gamma(t)$ observed already
in~\cite{Amon2012} is recovered, with a smaller slope in the case of
$h=5\ cm$ (fine orange line) compared to the case $h=0\ cm$ (fine
green line). When the stress modulation is turned on, we also obtain,
after a transient regime, a proportional relationship between
$N_{c}(t)$ and $\gamma(t)$ but with a smaller slope.  This
proportionnality shows that the macroscopic plastic deformation is the
result of the accumulation of local plastic events. The coefficient of
proportionality depends thus on the value of $h$ but also on the
regimes (relaxation or rectified-regime). For the latter, the
``hot-spots" production rate inducing similar plastic deformations
when compared to the relaxation-regime values, is significantly
smaller (Fig.~\ref{fig.NG}a). The change of the slope is visible in
both experiments at $h=0$ and $5\ cm$. Finally, two experiments with
stress modulation for each depth are superimposed, showing the
reproducibility of the measurements.

\section{Rheological model}
The observed linear creep was interpreted in~\cite{Pons2015} as a
secular drift, i.e. the accumulation of tiny effects over a very long
time. The ingredients necessary for this sub-threshold fluidization by
mechanical fluctuations to take place are the combination of shear
rejuvenation and memory effect. Such ingredients are taken into
account in the simplest mathematical way in a model put forwards by
Derec et al. \cite{Derec2001}, to render the macroscopic phenomenology
of aging complex fluids. This model was used successfully to interpret
experimental strain relaxation curves and rectified creep in previous
works \cite{Nguyen2011,Espindola2012,Pons2015}. A central parameter in
this model is a fluidity variable $f$ representing the rate of
visco-elastic relaxation for the shear stress $\sigma$ :

\begin{equation}
\label{eq.fluidity1}
\dot{\sigma} = G\dot{\gamma }-f\sigma 
\end{equation}
$G$ is the shear elastic modulus. The rheological complexity stems
from the dynamical equation undertaken by the fluidity variable. Many
forms were suggested by Derec et al. \cite{Derec2001}, but in the
context of granular creeping flow the experiment results point on a
simple expression :

\begin{equation}
\label{eq.fluidity2}
\dot{f}=-af^2+r\dot{\gamma}^2
\end{equation}
where dimensionless parameters $a$ and $r$ represent respectively
ageing and shear-induced rejuvenation processes. This model naturally
induces a logarithmic relaxation under constant shear and also a
dynamical threshold $\sigma_D=G\sqrt{a/r}$. In the context of granular
matter this threshold value should be proportional to the confining
pressure to get a Coulomb dynamical friction
coefficient. Experiments~\cite{Nguyen2011} seems to indicate that
under large shear, major reorganizations may occur in the packing and
one should not consider anymore $a$ and $r$ as stress independent
parameters. Recently, Pons et al. \cite{Pons2015} have shown
theoretically that the threshold of a visco-elastic fluid describes by
this model will be destroyed by vanishingly small stress fluctuations
around a bias by a secular effect. Consequently, below the threshold
($\sigma_0 < \sigma_D$), an effective viscosity is expected of the
form:
\begin{equation}
\eta = \frac{G \sigma_D}{\omega \delta}\sqrt{2 \left( 1 -
  \frac{\sigma_0^2}{\sigma_D^2}\right)}.\label{eq.eff_visc}
\end{equation}

In the following, we will consider the present model in its simple
form ($a$ and $r$ constant) to see if at least qualitatively the
salient experimental outcomes can be recovered.

The rheological equations (\ref{eq.fluidity1}) and (\ref{eq.fluidity2})
present rescaling parameters which are for stress $\sigma_D=G\sqrt{a/r
}$, for deformation $\gamma_0=\sigma_D/G$, for fluidity
$R_\sigma/\sigma_D $ and for time $\sigma_D/R_\sigma $.  Then the
dimensionless equations are as follows

\begin{equation}
\label{eq.fluidity1bis}
      \dot{\sigma} = \dot{\gamma }-f\sigma
\end{equation}
\begin{equation}
\label{eq.fluidity2bis}
      \dot{f}      = -a( f^2 - \dot{\gamma}^2)
\end{equation}
Note that in this dimensionless representation the only remaining
material parameter is $a$ and the only control parameters describing
the stress are the dimensionless mean stress and amplitude:
$\sigma_0/\sigma_D$ and $\delta/\sigma_D$. This system of equations
can then be solved numerically to reproduce the experimental
protocol. In presence of stress modulation, the resulting plastic
deformation exhibits a transient logarithmic response followed by a
linear regime (Fig.~\ref{fig.DWS}f). As we established before~\cite{Pons2015}, this is
qualitatively what is observed experimentally. The model allows thus
to recover the general experimental behaviour as far as global
deformation is concerned. 

Several works have shown experimentally a direct relation between the
rate of plastic events and the fluidity variable
$f$~\cite{Amon2012,Jop2012}. However, it is not clear a priori that
the mechanical driving would generate the same modes of plastic
relaxation in the packing in the presence of the mechanical
perturbations. And indeed, in fig.~\ref{fig.NG}a, one can identify a
systematic change of slope for the relation between the cumulated
number of ''hot-spots'' and the mean strain in the presence of stress
modulation. This experimental fact leads to a central question on the
relation between the rate of ``hot spots" production and the fluidity
parameter. This is the central object of the incoming discuccion.

\section{Spatial response}
Provided the assumption of a constant linear relationship between the
rate of hot-spots production and the fluidity, the local and global
distribution of hot-spot events can be computed. First, the radial
heterogeneity of the stress field is taken into account in the model,
i.e. $\sigma(r) = \sigma(R_0) R_0^2/r^2$. However, here we do not
account for the vertical heterogeneities as for example would be the
case for experiments with the vane buried.

Eqs.~(\ref{eq.fluidity1bis}) and~(\ref{eq.fluidity2bis}) are then
integrated numerically for each $r$ giving local values of the
fluidity $f(r)$ and strain $\gamma(r)$. Because of the cylindrical
geometry, the total rate of hot-spots occurrence is then expected to
be proportional to $\int_{R_0}^R f(r)r dr$. The cumulated number of
spot $N_c(t)$ can then be obtained numerically (within a
multiplicative constant) by integrating this rate over time: $N_{c}(t)
= \int_0^t \int_{R_0}^R f(\tau,r)r d\tau dr $. In parallel,
$\gamma(r)$ is integrated over $r$ in order to obtain the rotation
angle $\alpha(t)$ of the vane. Thus we can obtained numerically the
mean strain $\gamma(t)=\alpha(t) R_0/(R-R_0)$ corresponding to the one
experimentally measured. Those two quantities are compared to the
dimensionless time in Fig.~\ref{fig.DWS}e-f. In the case of
Fig.~\ref{fig.DWS}e no stress modulation is considered and we recover
the behaviour of the model when no spatial dependence is taken into
account~\cite{Nguyen2011,Amon2012}. In the case of
Fig.~\ref{fig.DWS}f, stress modulation are imposed from $t=0$ and in
this last case we observe a change of creep regime after a transient
regime, in agreement with the experimental graphs of
Fig.~\ref{fig.DWS}b and d. Finally, Fig.~\ref{fig.NG}b shows the
relationship between $N_{c}(t)$ and $\gamma(t)$ obtained numerically
in each case. This relationship is the same as the one observed in
experiments: the integral of the fluidity is proportional to the
plastic deformation and for the rectified creep, a strong reduction of
the slope is observed when the transient is terminated.

\begin{figure}[ht]
\includegraphics[width=0.95\linewidth]{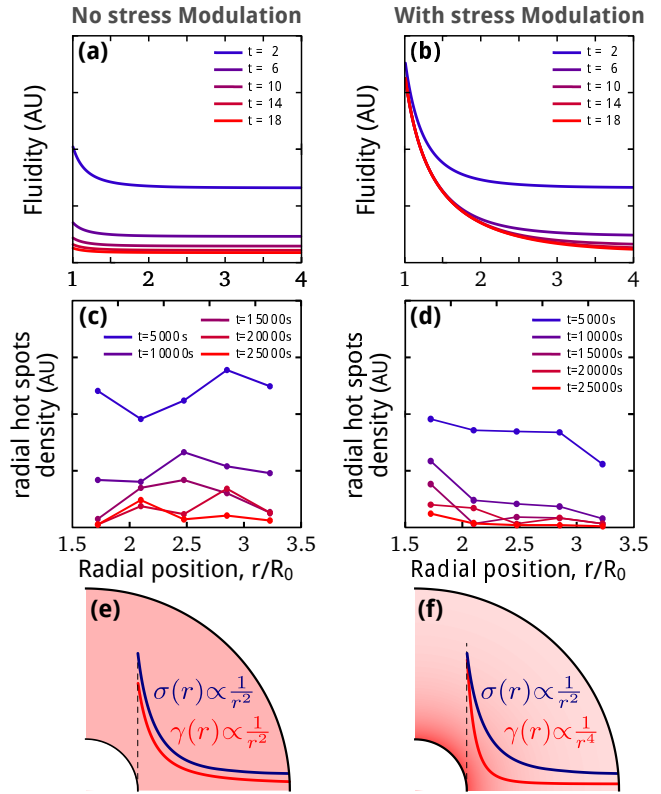}
\caption{(a-b) Numerical resolution of
  eq.~(\ref{eq.fluidity1bis}-\ref{eq.fluidity2bis}) taking into
  account the geometry of the system: evolution of the radial spatial
  fluidity over time whitout and with stress modulation,
  respectively. (c-d) Experimental results displaying the evolution of
  the radial hot spots density over time (experiments of
  Fig.~\ref{fig.DWS}c and d respectively). The vertical axis unit is
  arbitrary but identical for both experiments. (e) Schematic of the
  mechanical response in absence of modulation: the fluidity is
  spatially uniform in the cell (pink uniform background); the stress
  and the strain have the same spatial dependence. (f) Rectified case:
  the fluidity is non uniform and decreases as $\propto 1/r^2$.}
\label{fig.geometry_effect}
\end{figure}

Therefore, the outcome of the simulations is qualitatively very
similar to the experimental results. In particular, the change of
slope in Fig.~\ref{fig.NG}a is reproduced in the rheological model
when the heterogeneity of the stress field due to the geometry is
taken into account.  Microscopically, this reduction originates from a
difference in spatial distribution of ``hot spots" for the two cases
as we show numerically and experimentally in
Fig.~\ref{fig.geometry_effect}. Fig.~\ref{fig.geometry_effect}a
(resp. b) shows the radial evolution of the fluidity parameter without
(resp. with) stress modulation at different time of the numerical
integration. For the simple relaxation regime (logarithmic creep), the
fluidity is quite homogeneous across the cell and thus rather
independent of the local stress. The fluidity decreases uniformly with
time, reflecting the decrease of the strain rate. On the contrary, for
the rectified regime (Fig.~\ref{fig.geometry_effect}b), the fluidity
is quite different close or further away from the centre. It reaches
rapidly a permanent regime with a high density value close to the
centre while far from it, the density behaviour is close to the one
observed without modulation.

The radial distribution of hot-spots production can also be obtained
experimentally and and are shown in Fig.~\ref{fig.geometry_effect}c
(without modulations) and d (with modulations). The different curves
correspond to the time evolution of the density. The density is
calculated by averaging the number of ``hot spots" over a time windows
of $5000s$ and for a $5mm$ wide annulus. We first observe that this
density decreases with time for both experiments. Then, although the
data are quite noisy due to the small statistics, it seems that this
radial density stays homogeneous for the simple relaxation
(Fig.~\ref{fig.geometry_effect}c) while the apparition of ``hot spots"
decreases more rapidly far from the center of the cell during the
rectified creep (Fig.~\ref{fig.geometry_effect}d), which qualitatively
corroborates the outcome of the numerical measurements.

\section{Discussion}
To interpret this difference of behaviour in the radial spatial
repartition of fluidity in the two regimes, we use the analytical
results obtained from the perturbative analysis done
in~\cite{Pons2015}, but here, we explicitly take into account the
spatial dependence of the stress. In absence of modulation, the
fluidity is mostly governed by its initial value
$f_0$~\cite{Nguyen2011}:
\begin{equation}
\label{eq.fs_radial_na} 
f(r,t) = \frac{f_0}{1 + a f_0 \left( 1 -
  \frac{\sigma_0^2(r)}{\sigma_D^2}\right)t} \approx \frac{f_0}{1 + a f_0 t}
\end{equation}
Consequently, when $\sigma_0 \ll \sigma_D$, the non-rectified fluidity
is independent of the radial position, a result recovered both in
numerical solution (Fig.~\ref{fig.geometry_effect}a) and in experiment
(Fig.~\ref{fig.geometry_effect}c). Yet, it uniformly slowly decreases
with time. We thus obtain $\dot{\gamma}(r,t) = \frac{f(t)
  \sigma_0(r)}{G} \propto 1/r^2$. This case is schematically
represented in Fig.~\ref{fig.geometry_effect}e.

On the other hand, in the rectified case, as recalled in the
description of the rheological model (eq.~\ref{eq.eff_visc}), the
local stationary value $f^*(r)$ of the fluidity is for $\sigma_0 \ll
\sigma_D$~\cite{Pons2015}:
\begin{equation}
f^*(r) = \frac{\omega \delta(R_0)}{\sigma_D \sqrt{2}}
\frac{R_0^2}{r^2},\label{eq.fs_radial_a} 
\end{equation}
If we suppose that a stationary solution is reached across the cell,
we obtain for the local strain rate $\dot{\gamma}(r) = \frac{f^*(r)
  \sigma_0(r)}{G} \propto 1/r^4$. In this
case, represented in Fig.~\ref{fig.geometry_effect}f, the strain is
highly localized in the vicinity of the blades.

Because of the strong decrease of the fluidity with $r$ in the
rectified case, the total activity in the cell for a given mean strain
across the cell is smaller in the case of
Fig.~\ref{fig.geometry_effect}f than of
Fig.~\ref{fig.geometry_effect}e. This is the origin of the decrease of
the slope in Fig~\ref{fig.NG} in the rectified regime. Note that
further refinement of the discussion taking into account that the
stationary solution, in the rectified case, is not reached at large
$r$ does not change the picture. The duration of the transients are
nevertheless of major importance in such experiments and it can be
problematic to conclude on the nature of the creep. Indeed, the time
for the transient to reach the stationary solution $f^*$ is typically
$1/f^*$ which diverges when the targeted fluidity decreases. Actually,
the distinction between a rectified creep of very small $f^*$ and a
logarithmic creep could be pointless experimentally. The time to reach
the stationary solution may become increasingly long while
environmental background noise, always presents in practice, may
eventually hinder the latter.


\section{Conclusion}
In this report, we studied the creep response of a granular packing
below the Coulomb fluidization threshold, both in the case of a
logarithmic relaxation and for the mechanically rectified regime
leading to a linear creep. In both cases, the global plastic
deformation was monitored in parallel with the production rate of
local plastic events. The experimental results were compared to the
outcome of a simple visco-elastic model, solved numerically in the
cylindrical geometry, which associates the local fluidity parameter to
a rate of hot-spots production. Even though this model is quite
simple, the salient experimental features were reproduced
semi-quantitatively. First the effective linear relation between the
cumulated number of hot-spots and the plastic deformation is recovered
with a larger slope in the transient as compared to the rectified
regime. Second, the qualitative features of the spatial distribution
of hot spots are recovered in the model in both regimes: a weak radial
dependence and an almost uniform decrease with time in the logarithmic
creep case; and in the case of a linear creep, strong spatial
dependence of the stationary solution close to the inner cylinder.

Interestingly, at this point, we do not need any non-local model to
reproduce those generic experimental features. Heterogeneities of the
stress field need to be taken into account to fully understand the
data but not any spatial diffusion of the fluidity which is currently
taken into account in more sophisticated fluidity
models~\cite{Henann2013, Bouzid2013,Henann2014, Bouzid2015}. However,
it is important to note that this result does not necessarily exclude
the generic presence of non-local terms in the rheological picture. It
may simply mean that in the present experiments the non-local terms
would not contribute significantly to the rheology. The experiments
were performed at two different depths of the shearing vane, in the
bulk and at the surface with an over-load inducing an equivalent
confining ptressure. Hot-spots were observed in both configurations
but the slope between the cumulated number of hot-spots and the
plastic deformation differs. A further step to infirm or confirm that
a local model is sufficient for the interpretation of all our data
would be to test if this change of slope can be recovered when taking
into account the full stress spatial distribution when the vane is
buried or not. Indeed, the presence of hot-spots in the case when the
vane is buried may be indicative of the spatial propagation of the
plastic activity through the material bulk and thus of non-locality.

Finally, we must underline that most of the present results were
obtained rather far from the dynamical threshold, where essentially,
we could individualize the hot-spot apparition. Closer to the
threshold and possibly due to spatial coupling and avalanching events,
it would be interesting to see if quantitatively the simple local
description still holds.

\acknowledgments This work was funded by a CNES fundamental research
grant, a CNES post-doctoral grant and ANR Jamvibe.

\end{document}